\documentclass[8pt]{article}
\usepackage{arxiv}
\usepackage[utf8]{inputenc} 
\usepackage[T1]{fontenc}    
\usepackage{hyperref}       
\usepackage{url}            
\usepackage{booktabs}       
\usepackage{amsfonts}       
\usepackage{nicefrac}       
\usepackage{microtype}      
\usepackage{graphicx}		
\usepackage{amsmath}
\usepackage{amssymb}
\usepackage{widetext}
\usepackage[utf8]{inputenc}
\title{Reaching the sensitivity limit of a Sagnac gyroscope through linear regression analysis }
\author{
 Angela D.V.~Di~Virgilio, Umberto Giacomelli, Andrea Simonelli and Giuseppe Terreni\\
  INFN Sez. di Pisa,\\
   Polo Fibonacci, Largo B Pontecorvo 3, I-56127 Pisa, Italy \\
\texttt{angela.divirgilio@pi.infn.it} \\
\And
Andrea Basti, Nicol\`o Beverini, Giorgio Carelli, Donatella Ciampini, Francesco Fuso, Enrico Maccioni and Paolo Marsili\\
  Universit\`a di Pisa and INFN,\\ Dipartimento di Fisica "E. Fermi", Largo B Pontecorvo 3, I-56127 Pisa, Italy
       \And
       Carlo Altucci, Francesco Bajardi, Salvatore Capozziello, and Raffaele Velotta\\
       Universit\'a di Napoli and INFN,\\ Dipartimento di Fisica, Complesso Univ.\ Monte Sant'Angelo, via Cintia, Napoli, Italy
        \And
       Alberto Porzio\\
       CNR-SPIN and INFN, Napoli, Complesso Univ.\ Monte Sant'Angelo, via Cintia, Napoli, Italy
       \And
Antonello Ortolan\\
       INFN-National Lab. of Legnaro, viale dell'Universit\`a 2, I-35020, Legnaro (PD), Italy \\
}

\begin{document}
\maketitle
\maketitle
\begin{abstract}
The sensitivity to angular rotation of the top class Sagnac gyroscope GINGERINO is carefully investigated with standard statistical means, using 103 days of continuous operation and the available geodesic measurements of the Earth angular rotation rate. All features of the Earth rotation rate are correctly reproduced. The sensitivity of fractions of frad/s is attained for long term runs. This excellent sensitivity and stability put Sagnac gyroscopes at the forefront for fundamental physics, in particular for tests of general relativity and Lorentz violation, where the sensitivity plays the key role to provide reliable data for deeper theoretical investigations.  The achieved sensitivity overcomes the conventionally expected one for Sagnac ring laser gyroscopes.

\end{abstract}

\section{Introduction}
\label{intro}
Ring Laser Gyroscopes (RLGs) exploit the Sagnac effect -- i.e. the interference of two counter-propagating photon beams in a closed optical path -- to measure absolute rotation of the apparatus with respect to the local inertial frame \cite{Schreiber2013}. 
For the last 40 years large-scale versions of RLGs have been regarded as a promising tool for geodesy  and fundamental physics researches \cite{Ste,Scully}. 
A relevant contribution to geodesy is a more accurate estimate of Earth Rotation Parameters (ERP) \cite{G2019}, i.e. polar motion and sub-daily variations of universal time,  that correspond to direction and modulus of the Earth angular velocity vector, respectively. 
RLGs eventually should provide to the International Earth Rotation System (IERS) \cite{IERS1} a continuous, high--resolution measurement of the ERP, complementary to the well-established methods based on VLBI and GNSS (e.g. GPS, LSR, DORIS, etc.) data.
Further to ERP measurements with ground-based instrumentation, challenging tests of fundamental physics can be carried out with RLGs by studying the tiny residuals of the proper time difference between the counter-propagating photon beams, which last once any known rotation contribution (geophysical, geodetic or local) has been taken into account and subtracted. These residuals are directly connected to metric description of local space-time geometry \cite{CAPOZZIELLO2011167} or searches for Lorentz violations \cite{Jay2019}.\\
Unfortunately, the non--linearities induced by laser dynamics have made less attractive the RLG applications, and prevented the full comprehension of long term stability and sensitivity of RLGs.  The laser dynamics corrections \cite{Lamb,Aronowitz,Beghi,Cuccato,DiVirgilio2019,DiVirgilio2020} are unavoidable if  the RLG top sensitivity must be pushed towards testing space-time structures and symmetries, beyond the experimental results in gravitational or particle physics already available in the literature \cite{Will,AlanK}.
In this regard, we have demonstrated that effects of  non linear dynamics can be cancelled at a level of one part in $10^3$ and this was enough to push the sensitivity of our RLG to sub-prad/s rotation rates \cite{PRR}, paving the way to explore fundamental physics thanks to such an unprecedented sensitivity for a RLG. It is remarkable that also cold atoms in a Sagnac interferometer  achieved a pretty good sensitivity. However, there is a $\sim 5$ orders of magnitude gap in rotation sensitivity between current devices and large-scale RLGs \cite{CRPH,atomChina,atomFrance}. Another interesting gyroscope design is based on passive optical cavities which have been studied for gravitational waves detectors  \cite{Evans,Rana}, and are foreseen for space based gravitational waves antennas \cite{TQ,Jie1,Jie2}, but they have not yet demonstrated sufficient sensitivity.  

Notable General Relativity (GR) tests with gyroscopes are the measurement of the De Sitter effect due to the curvature of space-time around the Earth and of the Lense--Thirring effect due to the Earth rotation (dragging of the local inertial frames) \cite{PRD:2011}. Such tests are based on the comparison between the Earth angular velocity vector as estimated by IERS and the corresponding measurements obtained by an array of RLGs.  
Moreover, a RLG array can reconstruct the ``local geometry'' of  null geodetics of space-time and test whether it fully corresponds to the GR description or it requires GR extensions or modifications \cite{Aldrovandi:2013wha,Clifton:2011jh}.
Though at the level of the solar system GR well fits experimental observations, it suffers several shortcomings from the very small up to the cosmological scales. For example, it cannot predict the right correlation between mass and radius of some neutron stars \cite{Antoniadis:2013pzd, Demorest:2010bx}, the galaxy rotation curve without introducing Dark Matter \cite{Bosma:1981zz, Beringer:1900zz}, or the accelerated expansion of the Universe in the late time without introducing Dark Energy \cite{Copeland:2006wr}. Dark Matter and Dark Energy are supposed to represent the $26.8 \%$ and 68$\%$ of the Universe content, but have never been detected directly. 

At the small scales, while Strong and Electroweak interactions can be dealt with under the standard of quantum field theory, many issues arise in the attempt to merge the formalism of GR with that of quantum mechanics \cite{Ashtekar:2002sn, DeWitt:1967uc, tHooft:1974toh}. Indeed, in view of a possible quantum scheme, the spacetime metric should represent both a dynamical field and the background. This is not the case of other interactions, whose treatment is simplified by the assumption that the spacetime is supposed to be flat. Also, from quantum field theory in curved spacetime, a discrepancy of 120 orders of magnitude occurs between the theoretically predicted value of the cosmological constant and the experimentally observed one.

Modified theories of gravity arose with the purpose of solving such shortcomings, by considering alternatives to the Einstein--Hilbert action \cite{Capozziello:2019klx, Clifton:2011jh}. GR can be modified in several ways, such as introducing the coupling between geometry and scalar fields \cite{Bajardi:2020xfj, Elizalde:2004mq}, higher-order curvature invariants \cite{Bajardi:2020osh, Benetti:2018zhv, Blazquez-Salcedo:2017txk}, torsion and non-metricity \cite{Aldrovandi:2013wha, BeltranJimenez:2019tjy}, or by not requiring the equivalence principle to hold \emph{a priori} \cite{Bousso:2013ifa}.

In this context, experimental observations play a fundamental role. One the one hand, they can be used to constrain dynamical degrees of freedom occurring in modified theories of gravity \cite{Radicella:2014jwa}, selecting physically relevant theories. One the other hand, observations may address the research for GR extensions towards viable models, also suggesting the scales in which such extensions are needed.

For instance, in  \cite{Capozziello:2014mea}, post-newtonian approximation is used to put upper limits to the functional form of a higher-order scalar-tensor action; in \cite{Zakharov:2006uq} $f(R)$ gravity is constrained by solar system tests; in \cite{Capozziello:2017rvz} the fundamental plane of galaxies is addressed to geometric contributions; in \cite{Bahamonde:2017sdo} non-local theories of gravity are selected by S2 star orbit. 

Other interesting tests searching for new physics involve the local Lorentz invariance, since well motivated extensions of the Standard Model for particle physics predict Lorentz violation terms, that can be checked by the interference of the two counter-propagating beams of light \cite{Jay2019}. 

All such unique features of RLGs motivated us to propose the GINGER experiment \cite{Angela2017,Tartaglia2017a} and to build its test bed 
GINGERINO \cite{SIBelfi,90days} at the underground Gran Sasso laboratory.  GINGERINO is a 3.6 m side RLG that has been taking data in an almost continuous basis since 2017; it runs unattended and free running ensuring a duty--cycle around $80\%$ even in the absence of an active geometry control \cite{CQGGP2}.
This large amount of data gives us the possibility to improve the comprehension of the instrument, understand the sensitivity limits, and assess the feasibility of long term operation. Our activity is aimed at improving  the experimental set-up of the future GINGER array taking full advantage of the very careful analysis of GINGERINO data, where weak points of the system design can be identified, and  the origin of disturbances or  limitations in the sensitivity can be investigated. In general terms, the study of very high sensitivity apparatus is rather difficult, since it deals with noise. GINGERINO has the advantage to be sensitive to the global geodetic signals of the Earth, as Chandler and Annual wobbles, polar motions and variations of the universal time. These are rather small signals independently and constantly measured by the international system IERS with very high accuracy. Therefore, the analysis can effectively look for those signals in the GINGERINO data.

It has been recently demonstrated that GINGERINO reaches the sensitivity limit of 40 frad/s in 3.5 days of integration \cite{PRR}. The analysis has shown the dominant role of the tilt measurements in the identification and subtraction of the local disturbances. 
Aim of the present analysis is to improve the identification and subtraction of the local and instrumental disturbances by developing a more effective treatment of the signals produced by the tiltmeters implemented in the RLG set-up. To fully exploit the GINGERINO sensitivity we have also improved the cross calibration procedure with the IERS data. The approach enables attaining an extremely high sensitivity,  even better than one hundred times the Lense-Thirring effect with a  bandwidth corresponding to a 600 s integration time and long term operation.
Plan of the paper is as follows. In Sect. 2 the general scheme of the analysis is outlined. Sect. 3 reports the  results of the analysis applied to 103 days of data, illustrating the calibration procedure and the relative sensitivity, and indicating the main instrumental limits of the apparatus. In Sect. 4 we investigate the occurrence of signals due to deformation of the Earth crust. Sect. 5 reports a general discussion about the analysis and the GR tests. Conclusions are eventually drawn in Sect. 6.

\section{Purpose and general scheme of the analysis}
Purpose of the analysis is to reconstruct the Earth angular rotation rate and the instrumental disturbances using the RLG data and the environmental signals with the final goal of improving sensitivity and investigating its effective limit.
The procedure uses linear regression (LR) model \cite{Kay,Neter,Sen,direnzo} minimising the square of the difference between the evaluated and independently measured rotation rates. To this end it is mandatory to subtract the contributions induced by the non linear laser dynamics \cite{DiVirgilio2019,DiVirgilio2020}.
In the following the main properties of the Sagnac frequency, the Earth rotation rate, the experimental set-up, and the analysis components are summarised.

\subsection{The Sagnac frequency}
The Sagnac gyroscope is identified by the oriented area  $\mathbf{A}$ enclosed by the optical path and the perimeter $P$ corresponding to its length. The Sagnac beating signal $\omega_s$ \footnote{The  conventional symbol to denote the Sagnac signal is $f_s$, as it is practical to measure beating frequencies in Hz. We use instead $\omega_s = 2 \pi f_s$ because the angular frequency is used in back scattering calculations and its introduction leads to an adimensional scale factor in Eq.~1.} is proportional to the scalar product between $\mathbf{A}$ and  the total angular velocity  $\mathbf{\Omega}_T$ of the RLG optical cavity. Without loss of generality, we can write 
\begin{eqnarray}
    \omega_s =&2 \pi \frac{\mathbf{A}}{\lambda\, P}\cdot \mathbf{\Omega}_T\nonumber \\
    \mathbf{\Omega}_T =&  \mathbf{\Omega}_\oplus +  \mathbf{\Omega}_{loc}\\ 
    \textrm{SF} =& 2 \pi \frac{A}{\lambda\, P} \ \cos\gamma \nonumber \, 
\end{eqnarray}
where $\lambda$ is the  laser wavelength, and $\mathbf{\Omega}_{loc}$ is the sum of all possible angular velocities associated with  Earth crust deformations and instrument infinitesimal rotations. In general, local rotations are unknown in amplitude and direction but very small compared to the Earth rotation rate $\Omega_\oplus$, and so they contribute to the Sagnac signal as an additive perturbation. Here, $\mathbf{\Omega}_\oplus$ describes the Earth rotation rate and its orientation, $\gamma$ is the angle between the area vector and the rotation axis, corresponding to the laboratory co-latitude for horizontal RLGs, and SF is the scale factor. In our analysis the $\cos\gamma$ is associated with the scale factor to simplify the discussion, since effects of geometrical scale factor changes cannot be distinguished from orientation changes. The modulus and direction of $\mathbf{\Omega}_\oplus$ changes in time, however it is  continuously monitored by IERS.
In the following we will use uppercase $\Omega$ and lowercase $\omega$ for angular velocity (in units of rad/s) and  the corresponding Sagnac angular frequency, respectively\footnote{Note, however, that for the sake of clarity we will use frequency units, in Hz, when numerical evaluations of quantities and uncertainties related to angular frequencies $\omega$ will be given all through the text and figures.}  In addition, their time dependence will not be explicitly indicated. 
\subsection{Geodesic signals}
The international system IERS provides the data to describe  the Earth motion on a daily basis, from which it is possible to reconstruct the effect on GINGERINO, called geodesic signal $\Omega_{_{IERS}}$, as the sum of the average Earth rotation rate $\Omega_\oplus$, the Length of Day (LoD) changes, celestial pole offsets,    UT1 - UTC,    polar motion and diurnal and semi-diurnal variations produced by ocean tides \cite{IERS1}. In the present analysis the $\Omega_{_{IERS}}$ time series is  downloaded from Earth orientation center of the Paris observatory \cite{IERSD}. 
\subsection{GINGERINO experimental set up and data analysis}
GINGERINO is a RLG with a square laser cavity, 3.6m in side. It is installed horizontally with its area vector aligned with the local vertical. Its design is based on a hetero-lithic (HL) mechanical structure, the 4 mirrors at the corners of the cavity are contained inside vacuum tight boxes and connected together by vacuum tubes. 
Interested readers can find more details in the literature \cite{SIBelfi,90days}.
The mechanical structure is attached to a cross shaped monument made of granite, connected in the center to the underneath bedrock through a reinforced concrete block. The set-up is  located underground, where typical day-and-night temperature variations are strongly suppressed, and far from  anthropic disturbances. The apparatus is protected by a cabinet, far from the large experimental halls of the Gran Sasso laboratory, moreover the electronics is contained in a separated room. The laser optical cavity is aligned at the beginning of the run, and after that it operates continuously and unattended. The geometry is not electronically controlled, this implies that mode jumps and split lasing mode occur. Routinely more than $90\%$ of the data are of good quality \cite{90days}; however, for the present analysis data around laser mode jumps are discarded, leading us to keep  no more than $80\%$ of the complete data set. 
\subsection{Model and parameters  of the analysis}
The Sagnac frequency  $\omega_{s}$ has to be evaluated by taking into account laser dynamics contribution $\omega_{LD}$, (i.e. $\omega_{s} = \omega_{s0}-\omega_{LD}$,
where  $\omega_{s0}$ is a first estimation of the Sagnac signal, as described in details in Refs. ~\cite{DiVirgilio2019,DiVirgilio2020,PRR,Frontiers2020}), and then it can be  expressed as  
\begin{equation}
\omega_s  = K_{cal}\, \omega_{_{IERS}} + \omega_{loc} \, ,\\
\end{equation}
where $K_{cal}$ is a cross calibration constant, very close to 1,  found by comparing $\omega_s$ and $\omega_{_{IERS}}$, the angular frequency related to the $\Omega_{_{IERS}}$ data, and the term representing local rotations $\omega_{loc} = \omega_{env} + \omega_{ins}$, which takes into account signals of environmental and geophysical origin $\omega_{env}$, and rotations of instrumental origin $\omega_{ins}$. The present analysis does not distinguish between $\omega_{ins}$ and $\omega_{env}$, but they are kept separated as, in principle,  we could get rid of $\omega_{ins}$ with an improved design of RLG. To this aim, we are investigating  the causes of the disturbances of an instrumental origin with a dedicated analysis \cite{envpaper}.

The $\omega_{LD}$ contribution is evaluated  including in the LR the 6 explanatory variables that correspond to the signals collected at the output ports of the square cavity, i.e. the beat note of the two counter propagating beams, the DC amplitudes of the two mono beams $IS_{1,2}$,  
their AC amplitudes $PH_{1,2}$, and their relative phase $\epsilon$. Other explanatory variables come from the available environmental signals: temperature, pressure, air flow speed, the two channels ($\zeta_{1,2}$) of the tiltmeter located on top of the granite table. The procedure is iterative as some explanatory variables have to be elaborated using a rough estimate of $\omega_s$, $\omega_{loc}$ and some environmental signals.
\subsection{Explanatory variables of the linear regression (LR)}
\label{sec:1}
 The present analysis  extends and refines the multiple linear regression procedure outlined in our recent work \cite{PRR}. 
It is worth noticing that the environmental infinitesimal rotations affect at the same time the Sagnac angular frequency $\omega_s$ and the laser dynamics contribution $\omega_{LD}$. For this reason $\omega_{LD}$ is evaluated  at the beginning with  the complete set of available explanatory variables related to laser dynamics and environmental sources. In any case, the final  result remains unchanged if  all the terms are kept in the LR procedure until its end.
Outputs of the analysis are an estimate of the geodesic signal reconstructed from the RLG data $\omega_{geo}$, the local angular velocity $\omega_{loc}$, and  the residual of the model $\Delta M_{LR}$, which provides an insight into physical phenomena not included in the proposed model. 
 The term accounting for local rotations  and the residuals read, respectively, 
\begin{eqnarray}
 \omega_{loc} = \omega_{s0}  - \sum_{i=1}^N a_i \, \omega_{LDi} - K_{cal}\  \omega_{_{IERS}}   \\
    \Delta M_{LR} = \omega_{s0} -K_{cal}\omega_{_{IERS}}  - \sum_{i=1}^N a_i 
   \omega_{LDi}\\ - \sum_{j=1}^M b_j\, env_j - 
    \sum_{k=1}^{K} c_k\, \tilde{F}_{k} 
     \nonumber, 
\end{eqnarray}
where $env_j$ are temperature, pressure, air flow speed,  and the two tiltmeter  signals $\zeta_{1,2}$, the time series $\tilde{F}_{k}$ are the product of the $\zeta_{1,2}$, or the DC monobeam signals $PH_{1,2}$, mainly with  $\omega_{loc}$, but also with residuals of an intermediate stage, and $a_i$, $b_j$, $c_k$ are weights of the LR procedure. The use of the explanatory variables $\tilde{F}_k$ has been suggested by our recent work \cite{PRR}. However, $\zeta_{1,2}$ and the relative products with $\omega_s$ were used in Ref. \cite{PRR} as separated explanatory variables. The present analysis indicated that the effective variable was the combination of the two, in the form of $\zeta_{1,2} \omega_{loc}$. 
To obtain  $K_{cal}$ and $\omega_s$, the LR is iterated a few times using at the first step only $\omega_{LD}$ to provide a first evaluation of $\omega_s$. The calibration value $K_{cal}$ is determined by imposing that at a fixed time T$_0$, arbitrarily chosen, the intercept $I_{T_0}$ of the LR is close to zero, i.e. $K_{cal} = (\Omega_{geo}(T_0)+I_{T_0})/\Omega_{_{IERS}}(T_0)$; in the iterative estimation of $\omega_s$ this procedure is repeated. \footnote{To correctly evaluate $\Omega_{_{IERS}}$ it is required to measure both the angle $\theta$ of the area vector of the RLG with respect to the rotational axis and the geometrical scale factor GSF 
$= 2 \pi A/(\lambda P)$. 
With a single RLG, as in the case of GINGERINO, we can estimate only a combination of the two quantities. Due to local rotations, the orientation of our apparatus is not fixed in time at the level of its sensitivity. Therefore, the model  cross calibrates at the time $T_0$ and any change  recovered by first and second order expansions is reconstructed by the LR analysis using the information of the explanatory variables.
Assuming FS the scale factor for horizontal orientation, the effective scale factor at the cross calibration point is $K_{cal}\, \textrm{FS}$, and so  $\omega_{s}(T_0) - \omega_{loc}(T_0) = K_{cal} \,  \omega_{_{IERS}}(T_0)$.} 
 The estimation of the geodesic angular frequency $\omega_{geo}$ is given by 
\begin{equation}
\label{eq:res}
    \omega_{geo} = \omega_{s0}  - \sum_{i=1}^N a_i \, \omega_{LDi}- \sum_{j=1}^M b_j\, env_j - \sum_{n=1}^{K1} c_n\, \tilde{F}_{n}\;.
\end{equation}
We noticed that $K_{cal}$ depends mostly on the absolute orientation of the device. Being $\theta$ the colatitude, and assuming that the geometry of the laser cavity is stable in time because the temperature is rather stable and geometrical variations are consequently small, it is possible to estimate the inclination of the RLG at the cross calibration time, $\theta_{css} =  \arcsin({\sin{\theta}\cdot K_{cal}})$. 

Although the complete set of explanatory variables is used in the initial analysis, we eventually keep only the statistical independent ones that affect the residuals. Accordingly, we neglect in our final elaboration pressure and air flow speed signals, since they demonstrated a negligible effect. 

\section{Analysis results}
\begin{figure}
    \centering
    \includegraphics[scale=0.45]{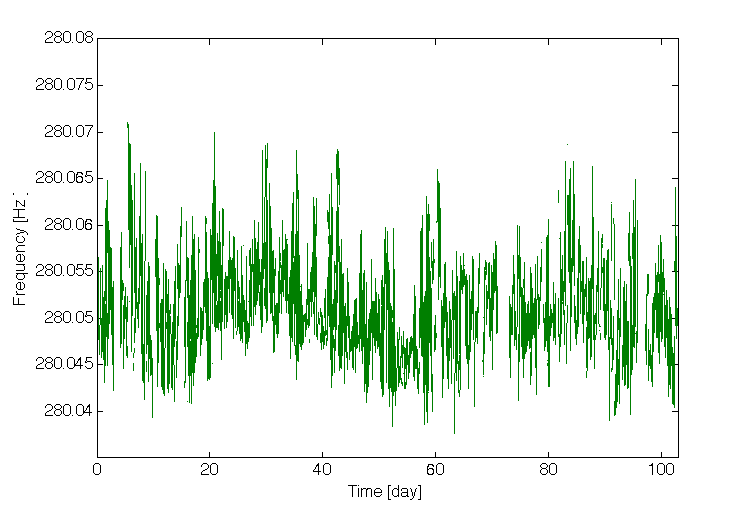}
    \caption{The evaluated Sagnac frequency $\omega_s/(2\pi)$.}
    \label{fig:true}
\end{figure}

\begin{figure}
    \centering
    \includegraphics[scale=0.30]{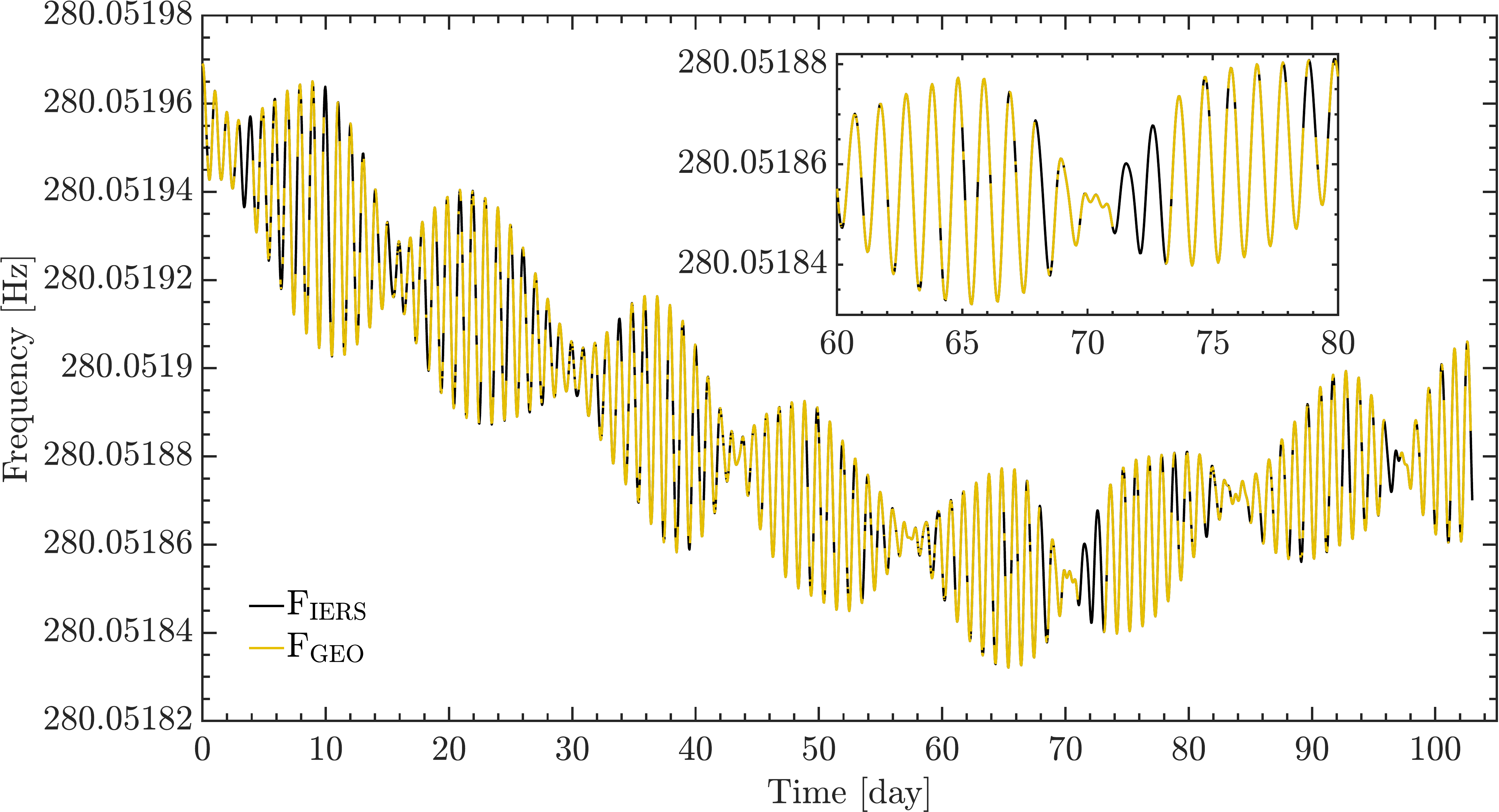}
    \caption{$\Omega_{geo}$ evaluated from the LR analysis applied to GINGERINO data compared with $K_{cal}\cdot\Omega_{IERS}$. The detail of 20 days is shown in the inset. The data removed from the analysis are evident.}
    \label{fig:geo}
\end{figure}

\begin{figure}
    \centering
    \includegraphics[scale=0.45]{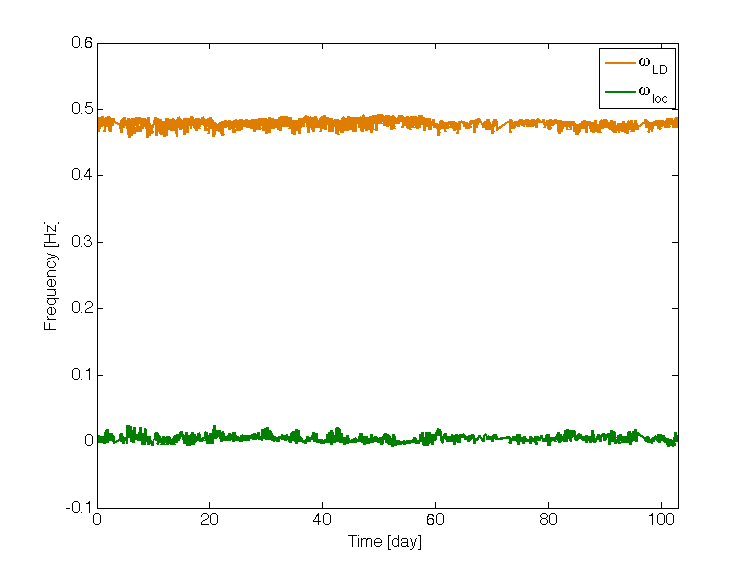}
    \caption{The total laser dynamics contribution $\omega_{LD}$ (orange) and local disturbances $\omega_{loc}$ (green) evaluated with the LR analysis applied to the complete time span.}
    \label{fig:LD}
\end{figure}

The LR method has been applied to a  set of data from day 1 to 103 of year 2020. The variables indicated in Eqs. 3-6 are time series with $\frac{1}{600}$Hz sampling rate. Typical results changing the cross calibration point reconstruct correctly $\Omega_{geo}$,  the angular velocity variations associated with $\omega_{geo}$. For example, at cross calibration $T_0$, modified Julian date 58886.23611, $K_{cal}= 1.000120452$, corresponding to a misalignment with respect to the RLG horizontal orientation of about 0.1 mrad (at the cross calibration time). We emphasize that  the relevant signals are reproduced  with a  standard deviation of the residuals $\Delta M_{LR}$ at the nHz level.
Figure \ref{fig:true} shows the evaluated true Sagnac signal $\omega_s$, and Fig. \ref{fig:geo} compares the effects on the RLG of $\Omega_{geo}$ and $\Omega_{IERS}$, expressed in units of frequency, Hz (corresponding in our notation to $\omega_{geo}/2\pi$ and $\omega_{_{IERS}}/2\pi$, respectively). The agreement between the IERS data and the evaluated $\Omega_{geo}$ is so good that the two curves are almost perfectly over imposed each other in Fig. \ref{fig:geo}.
Remarkably, the output of the LR analysis for the laser dynamics and local disturbances leads to non negligible contributions, as shown in Fig.~\ref{fig:LD}, reporting the evaluated $\omega_{LD}$ and $\omega_{loc}$, i.e.~the sum of the corresponding explanatory variables multiplied by the LR coefficients corresponding to laser dynamics and environmental monitor signals. Their standard deviations, in units of frequency, are as large as 6 mHz and 4.5 mHz, respectively, strongly suggesting that the related effects must be carefully accounted for in order to improve the instrumental sensitivity. Note that $\omega_{LD}$ has a bias close to 0.5  in units of frequency, remarking the importance of  the laser dynamics correction not only for the sensitivity, but as well for the accuracy, which is the key issue of GR tests.\cite{DiVirgilio2019,DiVirgilio2020}

The residual $\Delta M_{LR}$ is associated with portions of the data which cannot be explained by the model. Inclusion in the set of explanatory variables of those  created by multiplying $\omega_{loc}$ by $\zeta_{1,2}$ and the DC mono beam signals has impact in the results: we denote such variables as projectors, since they are scalar products. In particular, for GINGERINO, mainly  $\zeta_{1,2}$ signals are used in the projectors, being the use of the $PH_{1,2}$ less relevant. 
Typical $\Delta M_{LR}$ using a set of 4 or 5  projectors leads to a standard deviation as small as 4 nHz in units of frequency (corresponding to $0.7$ frad/s in angular velocity). 

\begin{figure}
    \centering
    \includegraphics[scale=0.28]{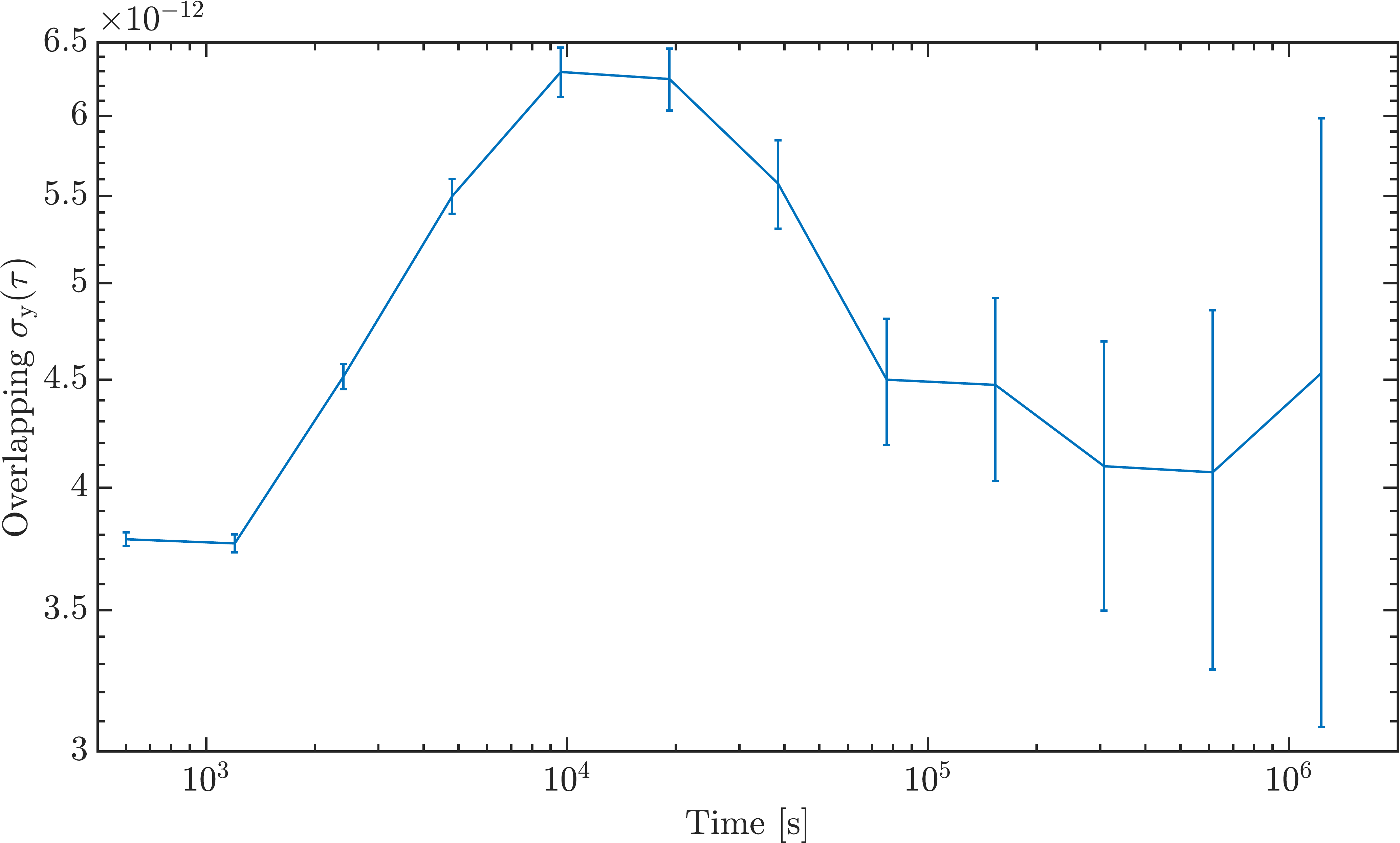}
    \caption{Typical Overlapping Allan Deviation of residuals $\Delta M_{LR}$, relative to mean $\omega_s$.}
    \label{fig:MAD}
\end{figure}

The Overlapping Allan Deviation (OAD) of the residuals $\Delta M_{LR}$, which provides information on noise and measurement sensitivity as a function of the integration time $\tau$, is shown in Fig. \ref{fig:MAD}. Remarkably, OAD is always below 6 parts in $10^{12}$ of the Earth rotation rate, well below the target meaningful for fundamental physics and geodesy, reported to be 1 part in $10^9$ \cite{Tartaglia2017a,Angela2017,PRR,Jay2019}. Moreover, the analysis has been repeated using 30 days of June 16-July 15 2018, obtaining a very similar behaviour of the OAD.

\subsection{Calibration by sinusoidal signal injection}
We implemented a supplementary method to estimate GINGERINO sensitivity by adding to $\omega_s$ two  sinusoidal signals. Then we run the LR procedure, including the explanatory variables that correspond to these probe signals. By looking at the results, in particular at the precision of the estimated probe parameters, we are able to calculate the signal to noise ratio and the sensitivity of the method at the frequency of the sinusoidal signals. In particular, the amplitude of the probes is  $2.6\cdot10^{-7}$ Hz (corresponding to the magnitude of the Lense-Thirring effect at the GINGERINO site), whereas their periods are 40 days and 0.5 days, respectively. They  are recovered by the LR analysis with a signal to noise ratio of $\sim 500$ and $\sim 1000$, respectively, corresponding approximately to noise floors of 0.5 nHz and 0.3 nHz in units of frequency (i.e. 0.1 frad/s and 0.05 frad/sec in angular velocity), with a  bandwidth corresponding to a 600 s integration time.

\subsection{Correlation between $\omega_{loc}$ and $\zeta_{1,2}$}
The analysis shows a strong correlation between RLG and tiltmeter signals.  Note that GINGERINO is a HL mechanical device, and its mirrors are not rigidly connected to the monument because there are mechanical levers used to align the square cavity. Those levers are not fixed and, in case of  monument tilts, the mechanical components will have different equilibrium configurations to compensate gravitational force. Since the mechanical parts are  connected to each other, the whole effect is an effective rotation of the laser cavity. From $\omega_{loc}$, it is possible to estimate the effective phase $\phi$ of the rotation. 
It has been straightforward to see a linear relation between the effective inclination of the monument and the reconstructed $\phi$ obtained by time integration of $\omega_{loc}$, the effective rotation of the device with respect to the ground.
This is a clear indication that the GINGERINO cavity rotates when the granite table changes its  orientation \cite{envpaper}. The correlation is not always linear, indicating that the HL mechanical cavity has a complex behaviour.

\section{Close look to the main features of the Earth rotation rate}
Known geophysical signals provide a real and effective playground for the analysis of GINGERINO data.
The Earth rotation rate contains several important features, as LoD effects, Earth normal modes and deformations induced by tides.  Accordingly $\Omega_{IERS}$ is the sum of different contributions, which can or not be taken into account in the analysis. Comparing the results obtained with different contributions to $\Omega_{_{IERS}}$ it is possible to isolate each contribution by subtracting the different $\Omega_{geo}$ elaborated by the different input models. In this way the analysis has been already able to reproduce LoD effects by providing the term $\Delta\omega_3$ \cite{PRR}. This procedure has been repeated to evaluate  LoD and the variations produced by ocean tides contributions, obtaining always results in agreement with the expectations, with a discrepancy even smaller than 1 part in $10^3$, consistent with the sensitivity of the apparatus, as shown for example in Fig. \ref{fig:MAD}.

\subsection{Effects of tides and Earth normal modes}
\begin{figure}
    \centering
    \includegraphics[scale=0.45]{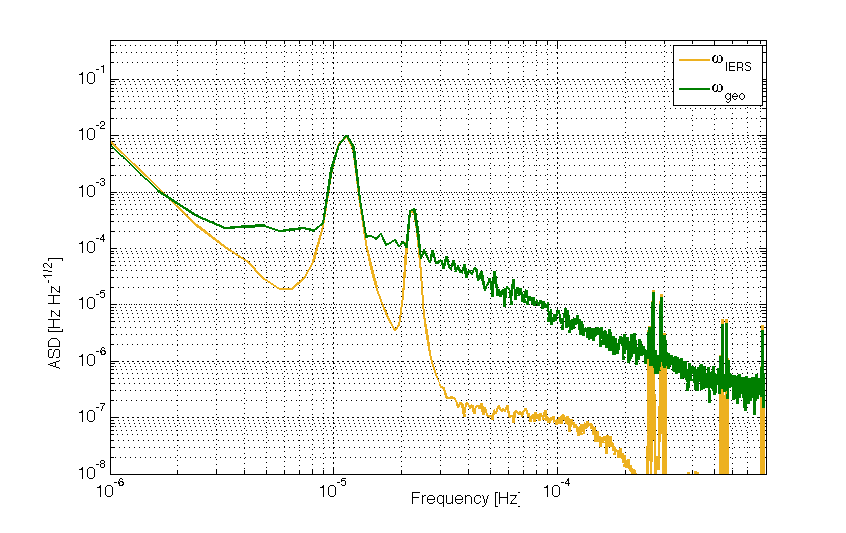}
    \caption{Comparison of the ASD of $\omega_{IERS}$ and $\omega_{geo}$: the main features due to polar motion and to solid Earth tide are clearly visible and well in agreement in both frequency and amplitude. Smaller and very tiny resonances at higher frequency are well visible. }
    \label{fig:SET}
\end{figure}

\begin{figure}
    \centering
    \includegraphics[scale=0.45]{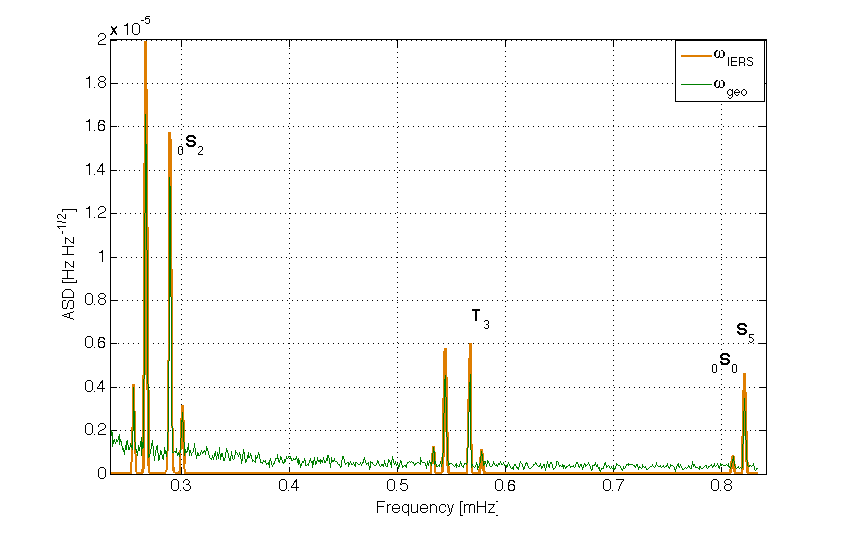}
    \caption{Detail of the rotation associated with Earth normal modes. The agreement with IERS expectation shows equal frequency, while the amplitude of the different components is systematically smaller by more than $10\%$. In the plot the positions of several well known resonances are qualitatively indicated for comparison. }
    \label{fig:NM}
\end{figure}

Since Earth is not a rigid body, tides and normal modes of the Earth induce crust deformations. As a consequence, angular rotations of the GINGERINO site occur; in the following we focus on the effects due to tides and normal modes of the Earth.
The angular rotation signal provided by IERS contains global signals, as polar motions, but also local signals caused by the deformation of the crust induced by tides or ocean loading. Figure \ref{fig:SET} compares the Amplitude Spectral Density (ASD) of $\omega_{IERS}$ and $\omega_{geo}$; the semi diurnal peak,  at a frequency slightly above $2\times 10^{-5}$ Hz, is caused by the solid Earth tide, $\omega_{geo}$ and $\omega_{IERS}$ are in good agreement each other in that frequency region.
Deformations associated with Earth normal modes have a typical frequency above  $0.3$ mHz (corresponding to a period below 53.9 prime minutes). Figure \ref{fig:NM} reports an expanded view of the ASD in the relevant frequency range. As demonstrated by the comparison of predictions and analysis results, very small and tiny peaks due to deformations are well reproduced, although peak amplitude is systematically smaller by more than $10\%$ with respect to the predictions based on IERS data.
Owing to the small amplitude of the signals relating to normal modes, much smaller than typical disturbances of the apparatus, at the present stage of the analysis we cannot definitely conclude that the instrumental sensitivity is large enough to truly reconstruct their occurrence in the considered frequency range.  In general the agreement is very good in the semi-diurnal signal, corresponding to a larger amplitude, while normal modes are found systematically smaller in amplitude. 

\subsection{Attempts to evaluate the rotation caused by deformation}
\begin{figure}
    \centering
 \includegraphics[scale=0.45]{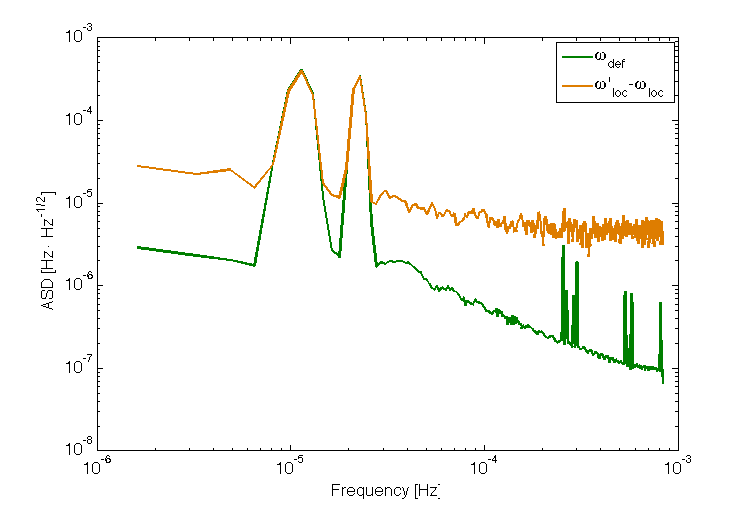}
 \caption{ASD of $\Omega_{def}$ as expected by IERS and evaluated with the present analysis.}
    \label{fig:def3}
\end{figure}

Further analysis has been done on this important subject, in order to recover the contribution  of the diurnal and semidiurnal variations produced by ocean tides, which will be called $\Omega_{def}$ in angular velocity and as $\omega_{def}$ in the relative effect on the Sagnac signal. The analysis pipeline has been repeated using $\Omega\sp{\prime}_{IERS}$, i.e. the expected global signal without the tide effects $\Omega_{def}$, as it can be obtained with the available online tools. For this analysis the June-July 2018 data set, 30 days, has been used, since at that time the temperature variations were a factor 5 smaller than for the 2020 data set, and the corresponding standard deviation of $\omega_{loc}$ was 4 times smaller. The estimation of $\omega_{def}$ is done subtracting the global and local signals estimated in the two different analysis.
In general, $\Omega_{geo}-\Omega\sp{\prime}_{geo}$ is equal to  $\Omega_{def}$ at the level of tens of frad/s, while the difference $\omega\sp{\prime}_{loc}-\omega_{loc}$  is close to $\omega_{def}$ with a noise of fractions of $\mu$Hz. 
Figure \ref{fig:def3} shows the amplitude spectral density of $\omega_{def}$ and $\omega\sp{\prime}_{loc}-\omega_{loc}$. In this case, the Earth normal modes are below the noise spectral density.

\section{Discussions and findings}
The analysis is based on the hypothesis that the  Sagnac frequency $\omega_s$ is the sum of different components: laser disturbances $\omega_{LD}$, geodesic global signals $\omega_{geo}$ (to be compared to the IERS measurement $\omega_{_{IERS}}$, which is used in the model) and local disturbances $\omega_{loc}$. The model attempts to estimate $\omega_{loc}$ as the sum of terms independently evaluated using the environmental signals as explanatory variables, in particular temperature, tilts and mono beam amplitudes. GINGERINO is affected by many disturbances, but since their amplitudes are very small, it is taken for granted that they can be eliminated from $\omega_s$ with the linear regression, assuming linear (or second order) expansion of the transfer functions of the environmental signals. It is worth noticing  that the $\omega_{LD}$ contribution is comparable to $\omega_{loc}$ and cannot be neglected, or in other words the LR with environmental signals is not able to identify the local disturbances. It is crucial to identify and subtract $\omega_{LD}$, since it prevents the estimation of the angular velocity of the apparatus  by means of a linear analysis approach, owing to the non linearities of the laser dynamics and the relevance of their effects. The described 
interpretation of GINGERINO data is confirmed by the comparison  with the monolithic RLG of the Wettzell Observatory G, which is not affected by large disturbances of instrumental origin, and  low frequency angular rotations around the vertical axis of geophysical origin\cite{KUlli}. 

The general outcome of our analysis is that $\omega_s$ is dominated by local disturbances, which can be eliminated below the 10 nHz noise level in units of frequency by using the tiltmeter data. A sensitivity of the order of the nHz, or even higher, can be reported. However,  the method relies on the accuracy of the explanatory variables and it cannot be considered predictive. In fact, we have 
the convergence of LR analysis (with larger residuals), even if we perturb some explanatory variables (for instance by slightly changing the Earth rotation model). \\
The analysis shows that the sensitivity of the apparatus is orders of magnitude better than expected  \cite{Schreiber2013}. 
Sensitivity is the key point for any fundamental physics application of RLGs, and this discrepancy with the expected noise level will deserve further investigation from theoretical side. On the one hand, it will be necessary to more carefully check the analysis procedure, in particular with more extended data sets, in order to verify whether part of the rotation signal is cancelled out. On the other hand, the development of a full Monte Carlo simulation of GINGERINO would create complete data sets, containing the laser dynamics and known angular rotation signals, to be subsequently reconstructed using the reported LR analysis. 

\subsection{Lesson learned on GR and Lorentz violation tests}

So far the measurements of the Lense--Thirring effect have been done by space experiments, using the gravity map of the Earth independently measured by the GRACE mission, and providing latitude averaged measurements \cite{ciuf:nature,GPB,lares2,Lucchesi_2015}. The promise of the GINGER project is the direct and local measurement of the main GR features of the rotating Earth, namely the de Sitter and Lense-Thirring effects. The sensitivity study indicates the feasibility of Lense-Thirring tests at the $0.1\%$ level, a factor 10 improvement with respect to the first GINGER proposal.  
Such a sensitivity level is also an important goal to discriminate among different theories of gravitation \cite{Capozziello2021}.
For the GR tests, high accuracy is necessary, and the cross calibration has to be replaced by independent measurements of the scale factor. This is feasible with accurate measurements of the geometrical scale factor and its electronic control \cite{CQGGP2}. At the same time, the inclination angle $\theta$ has to be evaluated independently. The scheme proposed in the GINGER project is to evaluate the relative angle using the RLG of the array oriented at the maximum Sagnac signal (area vector parallel to the north pole direction), which, being sensitive only at second order to local tilts, can be used as reference to evaluate the inclination angles with respect to the rotation axis of the other RLGs of the array. This is a crucial issue, since in this way the measurement of the inclination angle is limited only by the RLG noise. We note that the use of different schemes based on external metrology systems is in principle feasible. However, those external systems have to ensure continuous operation and angle accuracy at the prad level at least, otherwise the final sensitivity of the RLG array will be limited by the relative angle measurements.

Improvements to the HL mechanical scheme are suggested by our analysis, since it shows that local disturbances are mostly of an instrumental origin. This is also an important point for the capability of the analysis to determine differences with $\Omega_{_{IERS}}$. At present, local disturbances are hundreds of times larger than the signals we are looking for. Mechanical deformations and uncontrolled rocking of the HL set-up have to be reduced by improving the mechanical scheme. To this aim, specific tests can be carried out to measure the positions of the mirror holders with respect to the support structure as a function of changes of inclination and temperature. Long term thermal stability is also of paramount importance, see \cite{envpaper} for details.
Certainly the optimal experimental setup for testing gravitational theories is to operate two or more RLG arrays in separate sites. This setup is also advantageous for the other applications of RLGs in geophysics and geodesy, as enabled by the cross-disciplinary nature of the experiment.

The same remarks hold for high-sensitivity Lorentz violation tests with the advantage that the expected effect is modulated in time. Also in this case, the study of residuals $\Delta M_{LR}$  will provide hints for any new physics not modeled in the LR analysis \cite{Jay2019}.

\section{Conclusion}
GINGERINO is a top sensitivity RLG, running far from external disturbances and protected from large thermal excursions in the deep underground environment. Its data are compared with the global signal of the Earth rotation provided by IERS. The laser dynamics induces fluctuations of the Sagnac signals, which pose severe limitations to rotation sensitivity, due to their non linear nature. The analysis takes into account and eliminates the nonlinear laser disturbances and  recovers the global IERS signal with all its features. At the same time  disturbances of an instrumental and local origin are estimated.  The obtained residuals, i.e. the unmodeled   part of the Sagnac signal, with  a standard deviation of the order of a few nHz, indicate a rotational sensitivity below the frad/s level. Disturbances are mainly of an instrumental origin, suggesting the need for improvements in the mechanical design. By injecting probe signals in the GINGERINO data, we conclude that the sensitivity is $0.1$ frad/s with 600 s integration time, more than a factor one hundred below the expected noise for this class of instruments. 
In the near future it will be necessary to investigate this aspect from a theoretical side, and it will be necessary to develop a detailed Monte Carlo study, including the non linear dynamics of the laser, in order to carefully investigate the analysis procedure.  Once again we  remark that  GR tests require the RLG array of the GINGER project 
which, in principle, can provide independent measurements of the scale factors and relative angles\cite{Angela2017}. 

Work is in progress to apply this analysis scheme to the RLG of the geodesic observatory of Wettzell, to verify whether it is possible to remove laser dynamics fluctuations and improve its sensitivity. 
\section*{Acknowledgments}
We thank the Gran Sasso staff in support of the experiments, particularly Stefano Gazzana, Nazzareno Taborgna and Stefano Stalio .
We are thankful for technical assistance to   Alessio Sardelli and Alessandro Soldani of INFN Sezione di Pisa and Francesco Francesconi of Dipartimento di Fisica.
A special thank to Gaetano De Luca of Istituto Nazionale di Geofisica e Vulcanologia for regularly checking Gingerino operation status.


\end{document}